\RequirePackage[mathlines]{lineno}
\documentclass{nature}
\usepackage{graphicx}
\usepackage{verbatim}
\usepackage{color}
\usepackage{float}
\usepackage[normalem]{ulem}

\title{Measurement of interaction between antiprotons } 

\begin{document}

\maketitle


\begin{abstract}
One of the primary goals of nuclear physics is to understand the force between nucleons, 
which is a necessary step for understanding the structure of nuclei and how nuclei 
interact with each other. Rutherford discovered the atomic nucleus in 1911, and the large body of knowledge about the nuclear force since acquired was derived from studies made on nucleons or nuclei. Although antinuclei up to antihelium-4 have been discovered~\cite{AntiAlpha} and their masses measured, we have no direct  knowledge of the nuclear force between antinucleons. 
Here, we study antiproton pair correlations among data taken by the STAR experiment~\cite{STAR} at the Relativistic Heavy Ion Collider (RHIC)~\cite{RHIC} and show that the force between two antiprotons is attractive. In addition, we report two key parameters that characterize the corresponding strong interaction: namely, the scattering length ($f_0$) and effective range ($d_0$).
As direct information on the interaction between two antiprotons, one of the simplest systems of antinucleons, 
our result provides a fundamental ingredient for understanding the structure of more complex antinuclei and 
their properties.

\end{abstract}

Although the theory of Quantum Chromodynamics (QCD) provides us with an understanding of the foundation of the nuclear force, this binding interaction in nuclei operates at low energy, where the force is strong and difficult to calculate directly from the theory. For that reason, a common parameterization of the effective interaction between nucleons is based on experimental measurements, and the corresponding parameterization for antinucleons remains undetermined. The important parameters in such a description of the interaction are the scattering length ($f_0$), which is related to elastic cross sections, and the effective range of the interaction ($d_0$), which is determined to be a few femtometers (the typical nuclear scale). The existence and production rates of antinuclei offer indirect information about interactions between antinucleons, and also have relevance to the unexplained baryon asymmetry in the universe~\cite{RiottoAndTrodden}. Antinuclei produced to date include antiprotons, antideuterons, antitritons, antihelium-3, and the recently discovered antihypertriton and antihelium-4 (see ref~\cite{AntiAlpha} and references therein). 
The interaction between two antinucleons is the basic interaction that binds the antinucleons into antinuclei, and this has not been directly measured previously. Of equal importance, one aspect of the current measurement is a test of matter-antimatter symmetry, more formally known as CPT --- a fundamental symmetry of physical laws under the simultaneous transformations of charge conjugation (C), parity transformation (P) and time reversal (T). While various prior CPT tests~\cite{PDGBook} have been many orders of magnitude more precise than what is reported here, there is value in independently verifying each distinct prediction of CPT symmetry~\cite{PDGBook}.  

Ultra-relativistic nuclear collisions produce an energy density similar to that of the Universe microseconds after the Big Bang, and the high energy density creates a favourable environment for antimatter production. The abundantly produced antiprotons provide the opportunity to 
measure, for the first time, the parameters $f_0$ and $d_0$ of the strong nuclear force between antinucleons rather than nucleons. 

The technique used to probe the antiproton-antiproton interaction involves momentum correlations, and it resembles the space-time correlation technique used in Hanbury-Brown and Twiss (HBT) intensity interferometry. Since its invention for use in astronomy by Robert Hanbury-Brown and Richard Q. Twiss in the 1950's~\cite{HBT1}, the HBT technique has been adopted 
in many areas of physics, including the study of the quantum state of Bose-Einstein condensates~\cite{Schellekens05},
the correlation among electrons~\cite{Kiesel02}
and among atoms in cold Fermi gases~\cite{Rom06}. A Bose-Einstein enhancement in particle physics was first observed in the late 1950's as an enhanced number of pairs of identical pions produced with small opening angles (GGLP effect)~\cite{GGLP60}. Later on, Kopylov and Podgoretsky noted the common quantum statistics origin of HBT and GGLP effects, 
and, through a series of papers~\cite{Kop74,KP75,Pod89}, they devised the basics of the momentum correlation interferometry technique. In this technique, they introduced the correlation functions (CF's) as ratios of the momentum distributions of correlated and uncorrelated particles, $C({\bf p}_1,{\bf p}_2) = \frac{P({\bf p}_1,{\bf p}_2)}{P({\bf p}_1)P({\bf p}_2)}$ with $C=1$ for no correlations, suggested the so-called mixing technique to construct the uncorrelated distribution by using particles from different collisions (events), and formulated a simple relation of the CF's with the space-time structure of the particle emission region. As a result, the momentum correlation technique has been widely embraced by the nuclear physics community~\cite{GKW79,Led04, Boal90,Lisa05}.

\vspace{1cm}
\begin{figure}[H]       
\centering
\makebox[2cm]{\includegraphics[width=0.5 \textwidth]{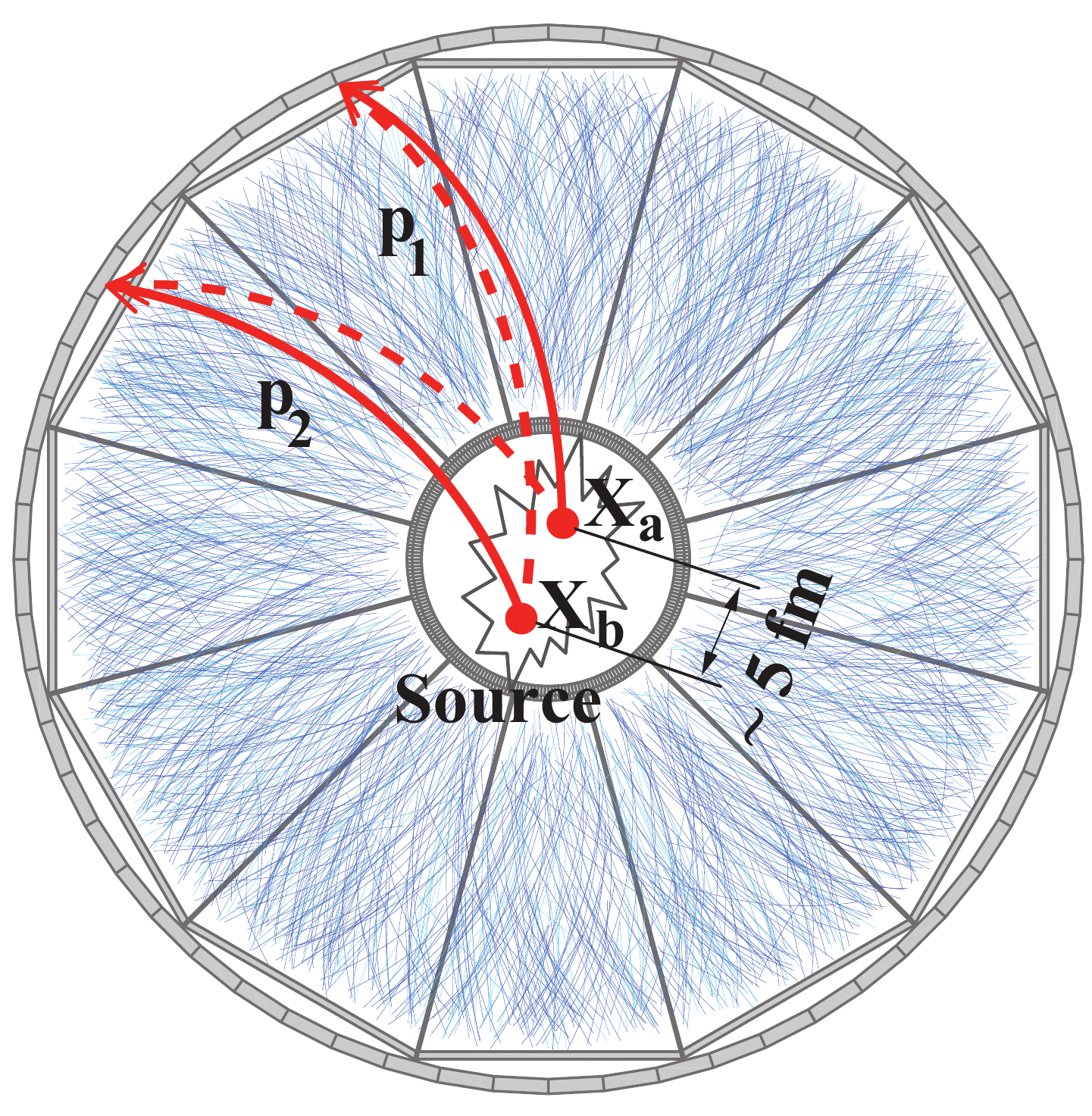}}
\caption{A schematic of the two-particle correlation process in a heavy-ion collision overlaid on an event display from the Time Projection Chamber in the STAR detector. The curves show particle trajectories, from which the track momenta are determined. These trajectories are 
measured in three dimensions, but are projected onto a single plane in this illustration. The STAR detector measures 3-vector momenta over a wide range beginning at about 0.1 GeV/$c$. Two particles emitted from two separated points, with four-coordinates $X_a$ and $X_b$, are detected with four-momenta $p_{1}$ and $p_{2}$. For the pair of indistinguishable particles with even/odd total spin, the two quantum mechanical amplitudes (representing, for non-interacting particles, products of plane waves 
$\langle p_1|X_a \rangle $ $\langle p_2|X_b \rangle $ and $\langle p_2|X_a \rangle $ $\langle p_1|X_b \rangle $, where $\langle p|X \rangle = {\rm exp}(-ipX)$)
 must be added/subtracted
 to yield the amplitude which is symmetric/antisymmetric with
 respect to the interchange of particle momenta.
 This results in an enhancement/suppression in the joint detection
 probability at zero momentum separation
 with the width inversely proportional to the space-time
 separation of particle emission points. }
  \label{fig:HBTCartoon}
\end{figure}        

Figure~\ref{fig:HBTCartoon} illustrates the process of constructing two-particle correlations in heavy-ion collisions.  In addition to quantum statistics effects, final state interactions (FSI) play an important role in the formation of correlations between particles. FSI include, but is not limited to, the formation of resonances, the Coulomb repulsion effect, and the nuclear interactions between two particles~\cite{Koo77,GKW79,Lednicky81, Boal90}. In fact, FSI effects provide valuable additional information. They allows for (see ref~\cite{Led04,RL08} and references therein) coalescence femtoscopy, correlation femtoscopy with non-identical particles, including access to the relative space-time production asymmetries, and a measurement of the strong interaction between specific particles. The latter measurement is  often difficult to access by other means and is the focus of this paper (for recent studies see ref~\cite{STARpLambda,STARLL}).

The data used here consists of Au + Au collisions at a centre-of-mass energy of 200 GeV per nucleon pair, taken during RHIC's operation in the year 2011. In total, 500 million events were taken by the minimum-bias trigger at STAR. The minimum-bias trigger selects all particle-producing collisions, regardless of the extent of overlap of the incident nuclei, but with a requirement that collisions must have occurred along the trajectory of the colliding Au ion and within $\pm 30$ cm of the center of STAR's Time Projection Chamber (TPC)~\cite{TPC}.  Events are categorized by their centrality, based on the observed number of tracks emitted from each collision. Zero percent centrality corresponds to exactly head-on collisions which produce the most tracks, while 100\%
centrality corresponds to barely glancing collisions, which produce the fewest tracks. Events used in this analysis correspond to the 30\%-80\% centrality class, for which the signal due to two-particle interaction is the clearest.

\begin{figure}[H]       
\centering
\makebox[2cm]{\includegraphics[width=0.5 \textwidth]{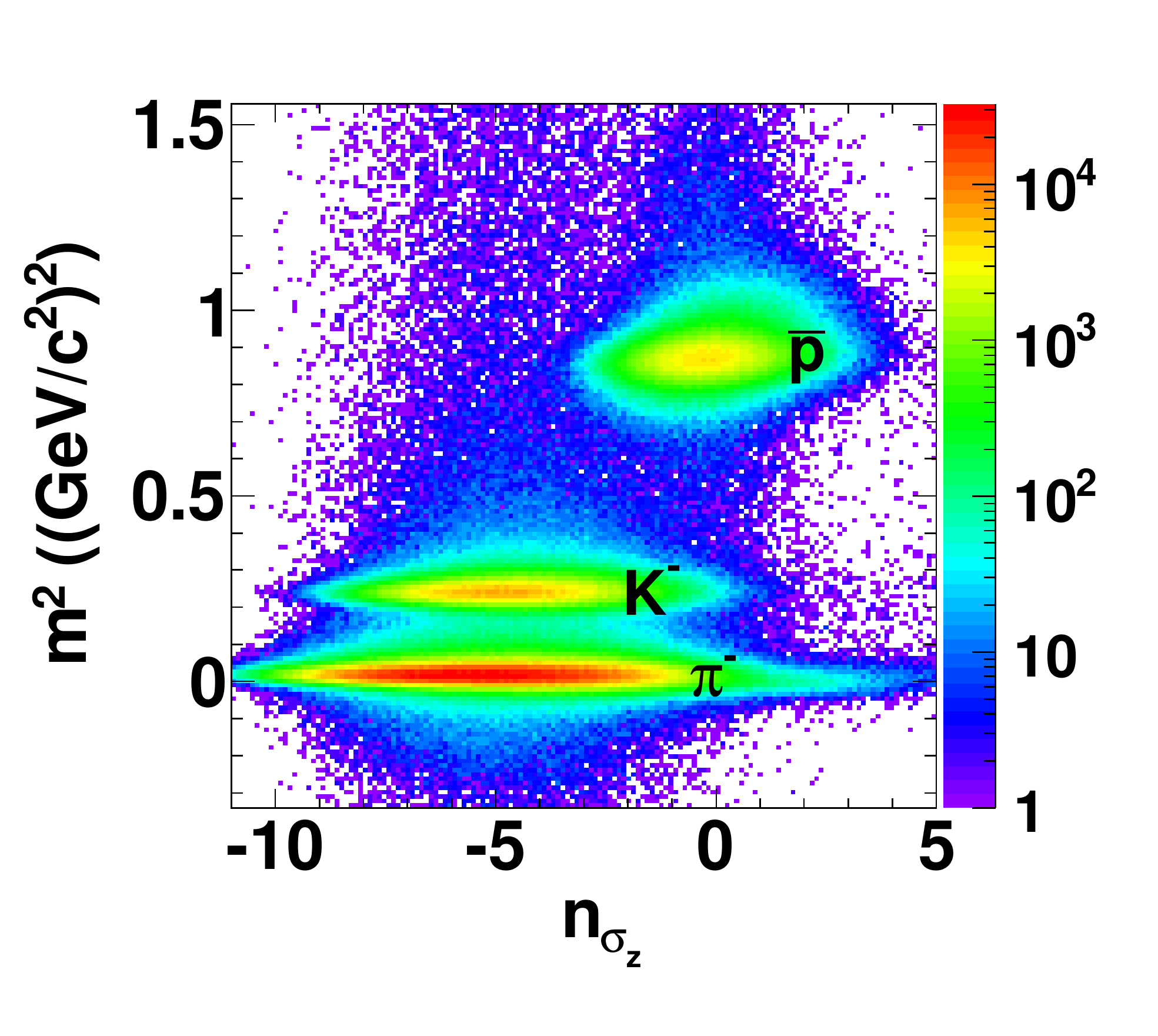}}
\caption{$m^2$ versus $n_{\sigma_{z}}$ for negatively charged particles. Here $m^2 = ({\bf p}^2/c^2)(t^2c^2/L^2 -1)$, where $t$ and $L$ are the time of flight and path length, respectively. $c$ is the light velocity. $z=\mathrm{ln}(\langle \mathrm{d}E/\mathrm{d}x \rangle / \langle \mathrm{d}E/\mathrm{d}x \rangle_E)$ and $\langle \mathrm{d}E/\mathrm{d}x \rangle_E$ is the expected value of $\langle \mathrm{d}E/\mathrm{d}x \rangle$ for (anti)protons. $\sigma_{z}$ is the r.m.s width of the $z$ distribution, and $n_{\sigma_{z}}$ is the number of standard deviations from zero, the expected value of $z$ for (anti)protons. The antiprotons, centered at $m^2=0.88 \, (\mathrm{GeV}/c^2)^2$  and $n_{\sigma_{z}}=0$, are well separated from other particle species for the momentum range specified. (Anti)protons satisfying $0.8\, {(\mathrm{GeV}/c^2})^2 < \mathrm{mass}^2 < 1\, {(\mathrm{GeV}/c^2})^2$  and $|n_{\sigma_{z}}| < 1.5 $ are selected for making pairs. With this selection, the purity is $>99\%$ for (anti)protons with transverse momentum less than 2 GeV/$c$.}
  \label{fig:PID}
\end{figure}

The two main detectors used in the measurement are the STAR TPC and the Time of Flight Barrel (TOF)~\cite{TOF}. The TPC is situated in a solenoidal magnetic field (0.5 T), and it provides a three-dimensional image of the ionization trails left along the path of charged particles. The TOF encloses the curved surface of the cylindrical TPC. In conjunction with the momentum measured via the track curvature in TPC, particle identification (PID) is achieved by two key measurements: the mean energy loss per unit track length, $\langle \mathrm{d}E/\mathrm{d}x \rangle$, which can be used to distinguish particles with different masses or charges, and the time of flight of particles reaching the TOF detector. Figure~\ref{fig:PID} shows a typical calculated mass-squared ($m^2$) distribution, derived from the time-of-flight and tracking information, versus $n_{\sigma_{z}}$ (see caption) for antiprotons. 

The population distribution of (anti)proton pairs as a function of (anti)proton momentum ($k^*$) in the pair rest frame is measured for the correlated pairs from within the same event, $A(k^*)$, and, separately, for the non-correlated pairs from two different (mixed) events, $B(k^*)$. The ratio of the two, $\frac{A(k^*)}{B(k^*)}$, gives the measured CF (see Methods). The observed (anti)protons can come from weak decays of already correlated primary particles, hence introducing residual correlations which contaminate the CF. The dominant contaminations to the CF come from the p-$\Lambda$ ($\bar{\mathrm{p}}$-$\bar{\Lambda}$) and $\Lambda$-$\Lambda$ ($\bar{\mathrm{\Lambda}}$-$\bar{\mathrm{\Lambda}}$) correlations, and are taken into account by fitting the CF with corresponding contributions. Taking the two-proton correlation measurement as an example~\cite{HannaThesis},  
\begin{eqnarray}
C_{\rm inclusive}(k^*) = 1 + x_{pp}[C_{pp}(k^*;R_{pp})-1]+x_{p\Lambda}[C_{p\Lambda}(k^*;R_{p\Lambda})-1]+x_{\Lambda\Lambda}[C_{\Lambda\Lambda}(k^*)-1], 
\label{eq:Inclusive}
\end{eqnarray}
where $C_{\rm inclusive}(k^*)$ is the inclusive CF, and $C_{pp}(k^*;R_{pp})$ is the true proton-proton CF, which can be described by the Lednick\'{y} and Lyuboshitz analytical model~\cite{Lednicky81}. In this model, for given s-wave scattering parameters, the correlation function with FSI is calculated as the square of the properly symmetrized wave function averaged over the total pair spin and the distribution of relative distances of particle emission points in the pair rest frame (see Methods). $C_{p\Lambda}(k^*;R_{p\Lambda})$ is the proton-$\Lambda$ CF from a theoretical calculation~\cite{Lednicky81}, which includes all final-state interactions and explains experimental data well~\cite{STARpLambda}. $C_{\Lambda\Lambda}(k^*)$ is from an experimental measurement corrected for misidentified $\Lambda$'s~\cite{STARLL}. $R_{pp}$ and $R_{p\Lambda}$, assumed numerically to be the same, are the invariant Gaussian radii~\cite{STARpLambda} from the proton-proton correlation and the proton-$\Lambda$ correlation, respectively. $x_{pp}$, $x_{p\Lambda}$ and $x_{\Lambda\Lambda}$, taken from the THERMINATOR2 model~\cite{THERMINATOR2}, are the relative contributions from pairs with both daughters from the primary collision, pairs with one daughter from the primary collision and the other one from a $\Lambda$ decay, and pairs with both daughters from a $\Lambda$ decay, respectively.

\begin{figure}[H]       
\centering
\makebox[2cm]{\includegraphics[width=0.5 \textwidth]{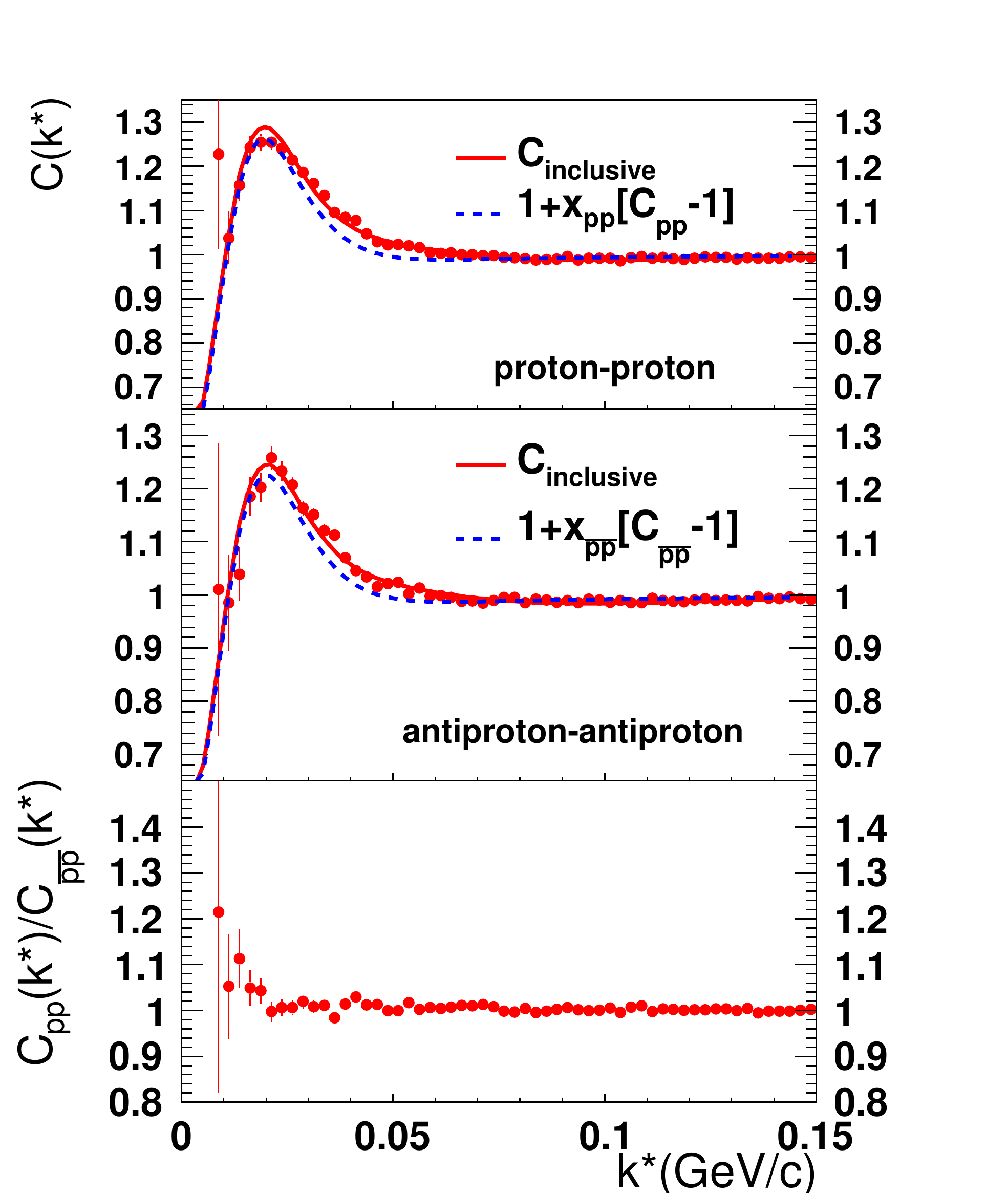}}
\caption{Correlation function for proton-proton pairs (top), antiproton-antiproton pairs (middle), and the ratio of the former to the latter (bottom). Errors are statistical only. The fits to the data with equation~\ref{eq:Inclusive}, $C_{\rm inclusive}(k^*)$, are plotted as solid lines, and the term $1 + x_{pp}[C_{pp}(k^*;R_{pp})-1]$ is shown as dashed lines. The $\chi^2/$ndf of the fit is 1.66 (1.61) for the top (middle) panel.  To take advantage of the existing knowledge on the proton-proton interaction, which is relatively well understood, when fitting the proton-proton correlation, $f_0$ and $d_0$ for protons are fixed at values measured from proton-proton elastic-scattering experiments, which are 7.82 fm and 2.78 fm, respectively~\cite{Mathelitsch1984}. When fitting the antiproton-antiproton correlation, $f_0$ and $d_0$ are treated as free parameters. }
  \label{fig:CF}
\end{figure}        
Figure~\ref{fig:CF} shows the CF for proton-proton pairs (top panel) and antiproton-antiproton pairs (middle panel), for the $30-80\%$ centrality class of Au + Au collisions at a centre-of-mass energy of 200 GeV per nucleon pair. The proton-proton CF exhibits a maximum at $k^*\approx 0.02$~$\mathrm{GeV}/c$ due to the attractive singlet s-wave interaction between the two detected protons and is consistent with previous measurements~\cite{Pochodzalla87}. The antiproton-antiproton CF shows a similar structure with the maximum appearing at the same $k^*$ value. In the bottom panel, the ratio of the inclusive CF for proton-proton pairs to that of antiproton-antiproton pairs is presented. It is well centered at unity for almost all the $k^*$ range, except for the region $k^* <$ 0.02 $\mathrm{GeV}/c$, where the error becomes large. This indicates that the strong interaction is indistinguishable within errors between proton-proton pairs and antiproton-antiproton pairs. By fitting the CF with equation~\ref{eq:Inclusive}, we determine the singlet s-wave scattering length and effective range 
for the antiproton-antiproton interaction to be $f_0=7.41 \pm 0.19 (\mathrm{stat}) \pm 0.36(\mathrm{sys})$ fm and $d_0=2.14 \pm 0.27 (\mathrm{stat}) \pm 1.34 (\mathrm{sys})$ fm, respectively.  The extracted radii for protons ($R_{pp}$) and that for antiprotons ($R_{\bar{p}\bar{p}}$), are $2.75 \pm 0.01(\mathrm{stat}) \pm 0.04(\mathrm{sys}) $ fm and $2.80 \pm 0.02(\mathrm{stat}) \pm 0.03(\mathrm{sys}) $ fm, respectively.

\begin{figure}[H]       
\centering
\makebox[2cm]{\includegraphics[width=0.5 \textwidth]{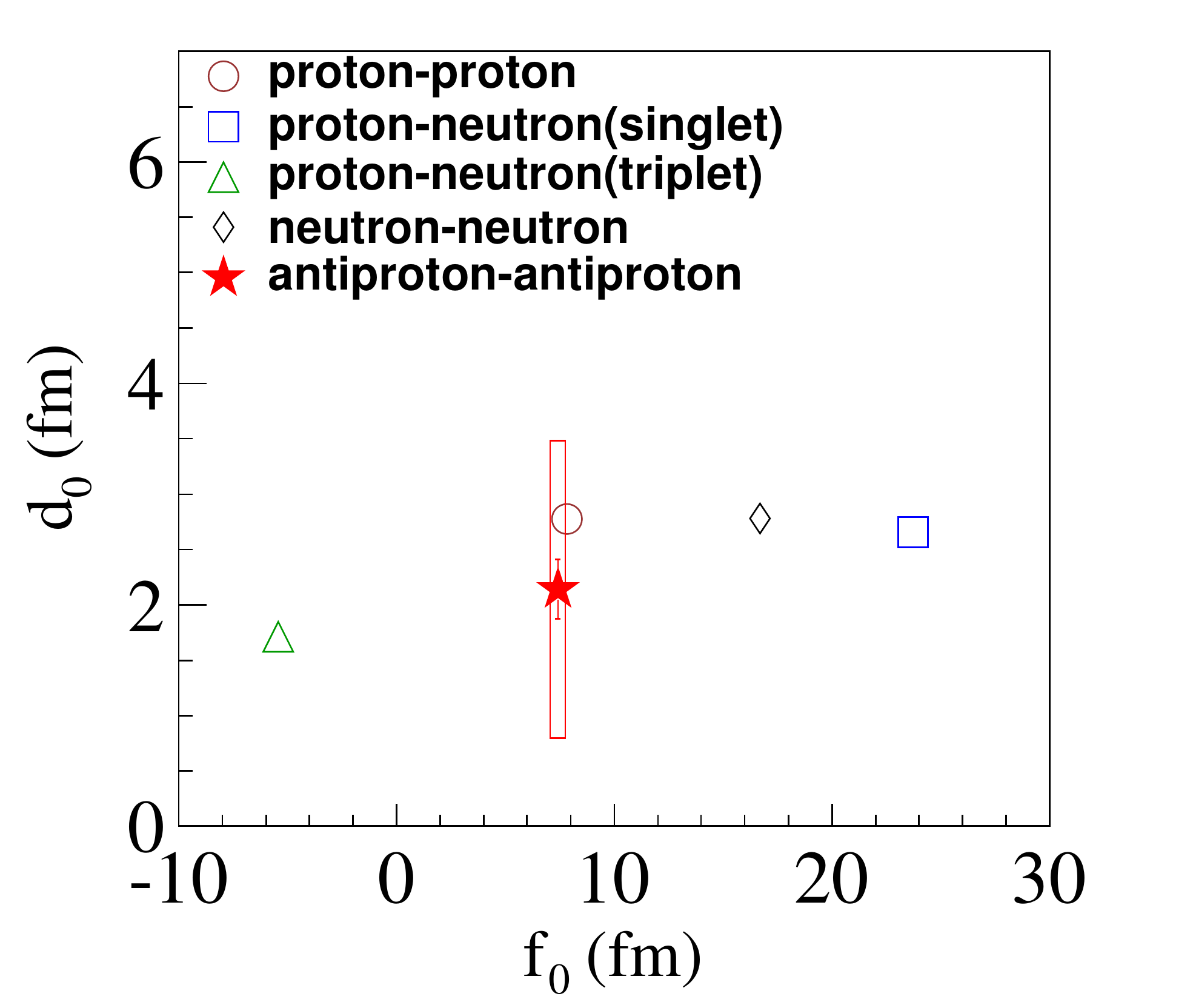}}
\caption{The singlet s-wave scattering length ($f_0$) and the effective range ($d_0$) for the antiproton-antiproton interaction plotted together with the s-wave scattering parameters for other nucleon-nucleon interactions. Here, statistical errors are represented by error bars, while the horizontal uncertainty for $f_0$ is smaller than the symbol size, and systematic errors are represented by the box. Errors on other measurements~\cite{Mathelitsch1984,Slaus89} are on the order of a few percent, smaller than the symbol size. }
  \label{fig:f0d0}
\end{figure}        

Figure~\ref{fig:f0d0} presents the first measurement of the antiproton-antiproton interaction, together with prior measurements for nucleon-nucleon interactions. Within errors, the $f_0$ and $d_0$ for the antiproton-antiproton interaction are consistent with their antiparticle counterparts -- the ones for the proton-proton interaction. Our measurements provide parameterization input for describing the interaction among cold-trapped gases of antimatter ions, as in an ultra-cold environment, where s-wave scattering dominates and effective-range theory shows that the scattering length and effective range are parameters that suffice to describe elastic collisions~\cite{Mott65}. The result provides a quantitative verification of matter-antimatter symmetry in the important and ubiquitous context of the forces responsible for the binding of (anti)nuclei.

\begin{methods}
\subsection{Event mixing for non-correlated pairs and the correction for purity.}
Non-correlated pairs each consist of two daughters particles. These daughters belong to two events which are  
carefully chosen so that they have similar event multiplicity and topology. The ratio of the $\frac{A(k^*)}{B(k^*)}$ (see above), after being normalized at a large $k^*$ (at least 0.25 $\mathrm{GeV}/c$), gives the measured CF, $C(k^*)_{\rm meas}$. 
Because in practice one cannot select 100\% pure (anti)protons, a correction to pairs is applied to obtain the PID-purity-corrected CF: $C_{\mathrm{PurityCorrected}}(k^*) = \frac{ C_{\mathrm{meas}}(k^*) -1}{\mathrm{PairPurity}(k^*)} + 1$. For simplicity, in Eq.~\ref{eq:Inclusive} the subscript ``meas" is dropped,  and elsewhere in this paper, the subscript ``PurityCorrected" is dropped.

\subsection{The transformation from $k^*_{p\Lambda}$ to $k^*_{pp}$.}
$C_{p\Lambda}(k^*)$ in equation~\ref{eq:Inclusive} is naturally expressed as a function of $k^*_{p\Lambda}$. Thus, to use it in the above equation, one needs to transform it to a function of $k^*_{pp}$. Here $k^*_{p\Lambda}$ (and $k^*_{pp}$) is the magnitude of the three-momentum of either particle in the pair rest frame, while in this case for $k^*_{pp}$, one of the protons is the decay daughter of $\Lambda$. This transformation is done by $C_{p\Lambda}(k^*_{pp})={\displaystyle \int} C_{p\Lambda}(k^*_{p\Lambda}) T(k^*_{p\Lambda},k^*_{pp}) dk^*_{p\Lambda} $, where $T(k^*_{p\Lambda},k^*_{pp})$ is a matrix that transforms $k^*_{p\Lambda}$ to $k^*_{pp}$~\cite{HannaThesis}. The transformation matrix is generated with the THERMINATOR2 model~\cite{THERMINATOR2} which is a Monte Carlo event generator dedicated to studies of the statistical production of particles in relativistic heavy-ion collisions.

\subsection{The calculation of the FSI contribution to the correlation function.}
The femtoscopic correlations due to the Coulomb FSI
between the emitted electron and the residual nucleus in beta
decay have been well known for more than 80 years;
they reveal themselves in a sensitivity of the Fermi function
(an analogue of the CF~\cite{Led07}) to the nuclear radius.
Compared with non-interacting particles, the FSI effect in a
two-particle system with total spin $S$ manifests itself in
the substitution of the product of plane waves, ${\rm exp}(-ip_1X_a-ip_2X_b)$,
by the non-symmetrized Bethe-Salpeter amplitudes
$\Psi_{p_1p_2}^{S(-)}(X_a,X_b) = \Psi_{p_1p_2}^{S(+)*}(X_a,X_b)$ ~\cite{GKW79,Lednicky81,Led09,Erazmus95}.
For identical particles, the symmetrization requirement in the
representation of total pair spin $S$ takes on the same form for
both bosons and fermions:
the non-symmetrized amplitude should be substituted by
$[\Psi_{p_1p_2}^{S(-)}(X_a,X_b) +
(-1)^S \Psi_{p_2p_1}^{S(-)}(X_a,X_b)]/\sqrt{2}$.
In the pair rest frame, $X_a-X_b=\{t^*,{\bf r}^*\}$
and $p_1-p_2=\{\omega_1^*-\omega_2^*,2{\bf k}^*\}$ where $\omega_i^*=(m_i^2+k^{*2})^{1/2}$ is the energy of
a particle of mass $m_i$, and $t^*$ and  ${\bf r}^*$ are the relative emission time and relative separation in the pair rest frame, respectively.
In this frame, the non-symmetrized  Bethe-Salpeter amplitude at equal emission
times ($t^*=0$) reduces, up to an inessential phase factor,
to a stationary solution of the scattering problem,
$\psi_{-{\bf k}^*}^{S(+)}({\bf r}^*)$.
At small relative momenta, $k^* <$~$\sim$$1/r^*$, this solution
can be used in practical calculations with the condition
$|t^*| \ll m r^{*2}$~\cite{Lednicky81,Led09}.
The equal-time approximation
is almost exact in beta decay, and it is usually quite
accurate for particles produced in high-energy collisions
(to a few percent in the FSI contribution to CF's of particles
even as light as pions~\cite{Led09}). In collisions involving heavy nuclei, the characteristic
separation of the emission points,  $r^*$, can be considered
substantially larger than the range of the strong-interaction
potential. The FSI contribution is then independent of the
actual potential form and can be calculated analytically
with the help of corresponding scattering amplitudes
only~\cite{gkll86}. At small $k^*$, it is basically determined by the s-wave
scattering amplitudes $f^S(k^*)$
scaled by the separation $r^*$~\cite{Lednicky81}.

\subsection{The analytical calculation of the (anti)proton-(anti)proton correlation function.}
The (anti)proton-(anti)proton correlation function, $C_{pp}(k^*;R_{pp})$ in equation~\ref{eq:Inclusive}, can be described by the Lednick\'{y} and Lyuboshitz analytical model~\cite{Lednicky81}. In this model, the correlation function is calculated as the square of the properly symmetrized wave function averaged over the total pair spin $S$ and the distribution of relative distances (${\bf r}^*$) of particle emission points in the 
pair rest frame, assuming 1/4 of the singlet and 3/4 of triplet
states and a simple Gaussian distribution
$dN/d^3{\bf r}^* \sim \exp(-{\bf r}^{*2}/4R_{pp}^2$).
Starting with the FSI weight of nucleons emitted with the
separation ${\bf r}^*$ and detected with the relative
momentum ${\bf k}^*$,
\begin{eqnarray*}
w({\bf k}^*,{\bf r}^*)=
|\psi_{-{\bf k}^*}^{S(+)}(({\bf r}^*) +
(-1)^S \psi_{{\bf k}^*}^{S(+)}({\bf r}^*)|^2/2,
\end{eqnarray*}
where $\psi_{-{\bf k}^*}^{S(+)}({\bf r}^*) $
is the equal-time ($t^*=0$) reduced Bethe-Salpeter amplitude
which can be approximated by the outer solution of the
scattering problem~\cite{Landau1974}. This is
\begin{eqnarray*}
\psi_{-{\bf k}^*}^{S(+)}(({\bf r}^*)=
e^{i\delta_c}\sqrt{A_c(\eta)}
[e^{-i{\bf k}^*{\bf k}^*}F(-i\eta,1,i\xi) +
f_c(k^*) \frac{\widetilde{G}(\rho,\eta)}{{r}^{*}}],
\end{eqnarray*}
where $\eta=(k^* a_c)^{-1}$, $a_c=$ (57.5 fm) is the
Bohr radius for two protons,
$\rho= k^* r^*$, $\xi= {\bf k^*}{\bf r^*}+\rho$,
$A_c(\eta)$ is the Coulomb penetration factor given by $A_c(\eta)=2\pi\eta[\exp(2\pi\eta)-1]^{-1}$,
$F$ is the confluent hypergeometric function,
$\widetilde{G}(\rho,\eta)=\sqrt{A_c(\eta)}[G_0(\rho,\eta)+iF_0(\rho,\eta)]$
is a combination of the regular ($F_0$) and singular ($G_0$)
s-wave Coulomb functions,
\begin{eqnarray*}
f_c(k^*)=[\frac{1}{f_0}+\frac{1}{2}d_0k^{*2}-
\frac{2}{a_c}h(\eta) - ik^*A_c(\eta)]^{-1}
\end{eqnarray*}
is the s-wave scattering amplitude renormalized by the Coulomb
interaction, and
$h(\eta)=\eta^2\sum\limits_{n=1}^{ \infty}[n(n^2+\eta^2)]^{-1}
-C -\ln|\eta|$
(here $C \doteq 0.5772$ is the Euler constant). The dependence of
the scattering parameters on the total pair spin S is omitted
since only the singlet ($S=0$) s-wave FSI contributes in the case
of identical nucleons.
The theoretical CF at a given $k^*$ can be calculated as the
average FSI weight
$\langle w({\bf k}^*,{\bf r}^*)\rangle$
obtained from the separation $r^*$, simulated
according to the Gaussian law, and the angle between
the vectors ${\bf k}^*$ and ${\bf r}^*$, simulated
according to a uniform cosine distribution.
This CF is subject to the integral correction~\cite{Lednicky81}
$-A_c(\eta)|f_c(k^*)|^2 d_0/(8\sqrt{\pi}R_{pp}^3)$
due to the deviation of the outer solution from the true
wave function in the inner potential region. In addition, in Au+Au collisions the emitting source has a net positive charge, and it influences the CF differently for proton and antiproton pairs. This effect is included in the consideration according to ref.~\cite{Led09,Erazmus95}

\subsection{Systematic uncertainties.}
The systematic uncertainties include variations due to track-wise and pair-wise cuts, the uncertainty in describing the $C_{p\Lambda}$ correlation function~\cite{BodmerAndUsmani}, and the uncertainty from the $C_{\Lambda\Lambda}$ measurement. The latter dominates the systematic error of $d_0$ and $f_0$, 
and it affects $d_0$ more than it does $f_0$ because the shape of the CF is sensitive to $d_0$, in particular at low $k^*$. As a consistency check, when fitting the proton-proton CF, both $f_0$ and $d_0$ are also allowed to vary freely, and the fitted $f_0$ and $d_0$ agree with the results from fitting the antiproton-antiproton CF. Assuming the measurements from different systematic checks follow a uniform distribution, the final systematic error is given by (maximum - minimum)/$\sqrt{12}$. In our calculations, we consider the two-proton wave function, taking into account the Coulomb interaction between point-like protons in all orbital angular momentum waves and the strong interaction in the s-wave only. We neglect the small non-Coulomb electromagnetic contributions due to magnetic interactions, vacuum polarization, and the finite proton size~\cite{Mathelitsch1984,Heller1967,Bergervoet1988}. This approximation changes the scattering parameters at the level of a few percent~\cite{Mathelitsch1984,Heller1967,Bergervoet1988}.

\end{methods}




\begin{addendum}
 \item We thank the RHIC Operations Group and RCF at
BNL, the NERSC Center at LBNL, the KISTI Center in
Korea, and the Open Science Grid consortium for providing
resources and support. This work was supported
in part by the Office of Nuclear Physics within the U.S.
DOE Office of Science, the U.S. NSF, the Ministry of Education
and Science of the Russian Federation, NNSFC,
the MoST of China (973 Program No. 2014CB845400),
CAS, MoST and MoE of China, the Korean Research
Foundation, GA and MSMT of the Czech Republic, FIAS
of Germany, DAE, DST, and UGC of India, the National
Science Centre of Poland, National Research Foundation,
the Ministry of Science, Education and Sports of the Republic
of Croatia, and RosAtom of Russia.
\end{addendum}

\author{
L.~Adamczyk$^{1}$,
J.~K.~Adkins$^{20}$,
G.~Agakishiev$^{18}$,
M.~M.~Aggarwal$^{30}$,
Z.~Ahammed$^{47}$,
I.~Alekseev$^{16}$,
J.~Alford$^{19}$,
A.~Aparin$^{18}$,
D.~Arkhipkin$^{3}$,
E.~C.~Aschenauer$^{3}$
% ...
}
\\
\\(STAR Collaboration) 
\\
\normalsize{$^{1}$AGH University of Science and Technology, Cracow 30-059, Poland}\\
\normalsize{$^{2}$Argonne National Laboratory, Argonne, Illinois 60439, USA}\\
\normalsize{$^{3}$Brookhaven National Laboratory, Upton, New York 11973, USA}\\
\normalsize{$^{4}$University of California, Berkeley, California 94720, USA}\\
\normalsize{$^{5}$University of California, Davis, California 95616, USA}\\
\normalsize{$^{6}$University of California, Los Angeles, California 90095, USA}\\
\normalsize{$^{7}$Central China Normal University (HZNU), Wuhan 430079, China}\\
\normalsize{$^{8}$University of Illinois at Chicago, Chicago, Illinois 60607, USA}\\
\normalsize{$^{9}$Creighton University, Omaha, Nebraska 68178, USA}\\
\normalsize{$^{10}$Czech Technical University in Prague, FNSPE, Prague, 115 19, Czech Republic}\\
\normalsize{$^{11}$Nuclear Physics Institute AS CR, 250 68 \v{R}e\v{z}/Prague, Czech Republic}\\
\normalsize{$^{12}$Frankfurt Institute for Advanced Studies FIAS, Frankfurt 60438, Germany}\\
\normalsize{$^{13}$Institute of Physics, Bhubaneswar 751005, India}\\
\normalsize{$^{14}$Indian Institute of Technology, Mumbai 400076, India}\\
\normalsize{$^{15}$Indiana University, Bloomington, Indiana 47408, USA}\\
\normalsize{$^{16}$Alikhanov Institute for Theoretical and Experimental Physics, Moscow 117218, Russia}\\
\normalsize{$^{17}$University of Jammu, Jammu 180001, India}\\
\normalsize{$^{18}$Joint Institute for Nuclear Research, Dubna, 141 980, Russia}\\
\normalsize{$^{19}$Kent State University, Kent, Ohio 44242, USA}\\
\normalsize{$^{20}$University of Kentucky, Lexington, Kentucky, 40506-0055, USA}\\
\normalsize{$^{21}$Korea Institute of Science and Technology Information, Daejeon 305-701, Korea}\\
\normalsize{$^{22}$Institute of Modern Physics, Lanzhou 730000, China}\\
\normalsize{$^{23}$Lawrence Berkeley National Laboratory, Berkeley, California 94720, USA}\\
\normalsize{$^{24}$Max-Planck-Institut fur Physik, Munich 80805, Germany}\\
\normalsize{$^{25}$Michigan State University, East Lansing, Michigan 48824, USA}\\
\normalsize{$^{26}$Moscow Engineering Physics Institute, Moscow 115409, Russia}\\
\normalsize{$^{27}$National Institute of Science Education and Research, Bhubaneswar 751005, India}\\
\normalsize{$^{28}$Ohio State University, Columbus, Ohio 43210, USA}\\
\normalsize{$^{29}$Institute of Nuclear Physics PAN, Cracow 31-342, Poland}\\
\normalsize{$^{30}$Panjab University, Chandigarh 160014, India}\\
\normalsize{$^{31}$Pennsylvania State University, University Park, Pennsylvania 16802, USA}\\
\normalsize{$^{32}$Institute of High Energy Physics, Protvino 142281, Russia}\\
\normalsize{$^{33}$Purdue University, West Lafayette, Indiana 47907, USA}\\
\normalsize{$^{34}$Pusan National University, Pusan 609735, Republic of Korea}\\
\normalsize{$^{35}$University of Rajasthan, Jaipur 302004, India}\\
\normalsize{$^{36}$Rice University, Houston, Texas 77251, USA}\\
\normalsize{$^{37}$University of Science and Technology of China, Hefei 230026, China}\\
\normalsize{$^{38}$Shandong University, Jinan, Shandong 250100, China}\\
\normalsize{$^{39}$Shanghai Institute of Applied Physics, Shanghai 201800, China}\\
\normalsize{$^{40}$Temple University, Philadelphia, Pennsylvania 19122, USA}\\
\normalsize{$^{41}$Texas A\&M University, College Station, Texas 77843, USA}\\
\normalsize{$^{42}$University of Texas, Austin, Texas 78712, USA}\\
\normalsize{$^{43}$University of Houston, Houston, Texas 77204, USA}\\
\normalsize{$^{44}$Tsinghua University, Beijing 100084, China}\\
\normalsize{$^{45}$United States Naval Academy, Annapolis, Maryland, 21402, USA}\\
\normalsize{$^{46}$Valparaiso University, Valparaiso, Indiana 46383, USA}\\
\normalsize{$^{47}$Variable Energy Cyclotron Centre, Kolkata 700064, India}\\
\normalsize{$^{48}$Warsaw University of Technology, Warsaw 00-661, Poland}\\
\normalsize{$^{49}$Wayne State University, Detroit, Michigan 48201, USA}\\
\normalsize{$^{50}$World Laboratory for Cosmology and Particle Physics (WLCAPP), Cairo 11571, Egypt}\\
\normalsize{$^{51}$Yale University, New Haven, Connecticut 06520, USA}\\
\normalsize{$^{52}$University of Zagreb, Zagreb, HR-10002, Croatia}\\


\end{document}